\documentstyle[11pt,preprint,aps]{revtex}
\input epsf
\epsfverbosetrue

\newlength{\oldbaselineskip}

\def\ee{e^{+}e^{-}}
\def\photino{\tilde{\gamma}}

\def\se{\widetilde e}
\def\ser{\se_{R}}

\def\eslash{E\hspace{-.09in}/}

\newcounter{figcount}

\draft
\preprint{TRI-PP-96-9}

\title{Supersymmetric Signatures at an $e$$\gamma$ Collider}
 
\author{Ken Kiers$^b$, John N. Ng$^{a,b}$ and Guohong Wu$^a$}

\address{$^a$TRIUMF Theory Group\\4004 Wesbrook Mall, Vancouver, B.C., 
Canada V6T 2A3\\ {\rm and} \\$^b$Department of Physics, 
University of British Columbia\\Vancouver, B.C., Canada V6T 1Z1}

\begin{document}
\setlength{\baselineskip}{24pt}
\setlength{\oldbaselineskip}{\baselineskip}
\maketitle

\begin{abstract}

\addtolength{\baselineskip}{-.3\baselineskip}
  High energy electron-photon colliders  provide unique 
 opportunities for probing physics beyond the standard model.
We have studied the experimental signatures for two  supersymmetric
scenarios, with the lightest supersymmetric particle (LSP) being
either the lightest neutralino or the gravitino.
In the ``neutralino LSP'' scenario favored by the minimal supersymmetric
standard model (MSSM), it is found that
 some basic parameters of the model, $\mu$, $\tan\beta$, $M_1$ and $M_2$,
may be uniquely determined from 
the outgoing electron energy spectrum without assuming high
scale unification of the masses or couplings.
In the ``gravitino LSP'' scenario which occurs naturally in models
of low energy dynamical supersymmetry breaking, 
it is possible to have
background-free signatures if the next-to-lightest supersymmetric particle
(NLSP) has a long decay length.
In cases that the NLSP decays quickly, ways to distinguish among 
the experimental signatures of the two scenarios and of the standard model 
(SM) background are discussed. 
\end{abstract}.

\setlength{\baselineskip}{\oldbaselineskip}

\newpage

\section{Introduction}
 
  It was first pointed out by Ginzburg, {\em et al}\cite{ginzburg1} 
that an $\ee$ collider could be adapted to make an $e\gamma$ collider 
by back-scattering laser light off the high energy electron beam.  
The resulting $e\gamma$ collider would be expected to have a 
center of mass energy and luminosity comparable to that of the $\ee$
collider, and this would open up new avenues \cite{egammaSM}
in testing the standard model (SM) of particle physics and 
in discovering physics beyond the standard model.

Experimental signatures of supersymmetry \cite{Niles,HK} at high energy 
$e\gamma$ colliders have been investigated to some extent 
\cite{preegamma,Cuy}. 
Assuming $R$-parity conservation, the lightest supersymmetric particle (LSP) 
is a stable neutral particle  and will escape detection.
The studies to date have focused on the simplest case
of lightest neutralino ($\chi_1$) production under the assumption that
$\chi_1$ is the LSP and is gaugino-like.
The process is $e\gamma \rightarrow \se \chi_1$ with subsequent
decay of the selectron $\se \rightarrow e\chi_1$,
and the experimental signature is $e + \eslash$.
A well-known feature of this process is that it has a lower threshold 
for selectron production than the analogous $\ee$ process.

  In this letter we extend the previous supersymmetric (SUSY) studies 
with $e\gamma$ colliders in two respects. 
Firstly, we will consider two scenarios, allowing the LSP
to be either the lightest neutralino $\chi_1$ or the gravitino.
The ``neutralino LSP'' scenario is often assumed to be realized
in the minimal supersymmetric standard model (MSSM) \cite{HK}, whereas in 
models of low energy dynamical supersymmetry breaking \cite{dine} 
the LSP is the gravitino and the next-to-lightest supersymmetric 
particle (NLSP) may be the lightest neutralino or a right-handed slepton,
whose decay proceeds  through its coupling to 
the Goldstino component of the gravitino \cite{Fayet}.  
Some of the phenomenology of a light gravitino has recently been discussed
in the context of $\ee$ and hadron colliders \cite{gravitinoph,kane}.
Secondly, we will generalize the previous analysis \cite{preegamma,Cuy}
in the MSSM (without assuming high scale unification of masses or couplings) 
by allowing for more production channels and decay modes of the $\se$.
We find that a complete determination of those MSSM parameters
that enter the neutralino and chargino mass matrices,
 $\mu$, $\tan\beta$, $M_1$, and $M_2$ (as well as the selectron mass), 
may be possible for certain ranges of these parameters by simply measuring 
the outgoing electron energy spectrum.

   It is well known that the cross section for 
$e\gamma \rightarrow W \nu \rightarrow e \nu \bar{\nu}$ approaches
a constant at asymptotic energies \cite{Ginz2}; 
hence it is crucial to reduce the $W$ background as much as possible. 
This can be achieved in two ways.
 The obvious one is to use highly-polarized right-handed electron beams.
At high energies
the electrons from the $W$ decay are peaked in the backward direction
inside the cone $\theta \sim m_W/\sqrt{s}$ ($\sqrt{s}$ is the center of mass
energy), so that the $W$ effects can 
also be significantly cut off by imposing angular cuts in 
the backward direction of the outgoing electron.
It is also expected that the photon beam obtained by back-scattering
laser light off the electron beam will not be monochromatic in practice.
   To simplify our analysis and focus on physics issues, 
we only consider the case with right-handed polarized electrons and
monochromatic high energy photon beams.
   Another advantage of using $e_R$ beams lies in the fact that 
the coupling of a right-handed electron/selectron to a chargino 
is suppressed by the electron mass, and this greatly simplifies 
the analysis.  
Furthermore, in $e_R\gamma$ collisions the SM backgrounds to the
SUSY signatures  most frequently involve a neutrino pair from 
$Z$ decay carrying away the missing energy, and this can be efficiently
eliminated by cuts on the invariant mass of the missing energy
around $m_Z$. 
Contamination from $e_L$ is expected in actual experiments and
can be easily incorporated.
Similarly, the folding in of the photon spectrum can be done once its actual
shape is known.

\section{Neutralino LSP Scenario}

   Within the minimal supersymmetric standard model,
the lightest neutralino state $\chi_1$  is often assumed to be the LSP. 
The mass of $\chi_1$ is generally expected to be at least 
tens of GeV \cite{pdg}.
We refer to the MSSM as the model with minimal particle content,
with $R$-parity conservation, and without
assuming gauge coupling, gaugino mass, or scalar mass unification at
high energy scales \cite{Niles,HK}.
In general, the states of the neutralino weak interaction basis 
mix among themselves and
 can be transformed into mass eigenstates ($\chi_i$, $i=1,2,3,4$)
 by a unitary matrix $N^{'}$ which is defined by the following
relation among two-component spinors
\equation
	\chi_i = N^{'}_{ij} \psi_{j}^{'} \label{eq:N},
\endequation
where $\psi_{j}^{'}$ contains the four states in the weak interaction
basis with $j=1,2,3,4$ referring to the photino ($\photino$), 
zino ($\widetilde{Z}$) and two orthogonal linear combinations of the
two higgsinos, respectively \cite{HK}.
Both the masses of the neutralinos and
the matrix $N^{'}$ are functions of the Higgs mixing parameter
$\mu$, the ratio of the two Higgs' VEVs $\tan \beta$, and the $U(1)_Y$ and
$SU(2)_L$ gaugino masses $M_1$ and $M_2$. 

\subsection{The Step Function Behavior of the Outgoing Electron Energy 
Spectrum}

  We start by considering the simplest case \cite{preegamma,Cuy} where
\begin{equation}
e_R\gamma \rightarrow \ser \chi_1 \;\;\;\;\;\;\;\;\;\;\;\;\; 
\mbox{and} \;\;\;\;\;\;\;\;\;\;\;\;\;
\ser \rightarrow e \chi_1 ,
\end{equation}
 assuming that $\chi_1$ is gaugino-like.
The experimental signature in this case is $e+\eslash$.
Throughout this letter we will use the narrow width approximation
which is justified because the sparticles decay weakly.  
In this approximation the differential cross sections 
can be easily obtained.
As the $\ser$ decay is isotropic in its rest frame, boosting back to
the lab frame (i.e. the $e\gamma$ CM frame) gives a flat distribution in 
the electron energy
spectrum\footnote{Folding in the photon energy spectrum will in general 
smear out the step function. However there are still an upper and a lower
bound on the outgoing electron energy which can be used to determine 
$m_{\se}$ and $m_{\chi_1}$ \cite{Cuy}.}.
The standard model background $e\gamma \rightarrow eZ$
with $Z \rightarrow \nu\bar{\nu}$
has an electron energy distribution with a Breit-Wigner shape, and can 
be eliminated by cuts on the outgoing electron energy. 

   The endpoints ($E_{\pm}$) in the electron energy spectrum can be
obtained from kinematical considerations. Inverting these relations
gives the masses of the $\ser$ and $\chi_1$,
\equation
	m_{\se}  =  \frac{s\sqrt{E_{+}E_{-}}}
		{\sqrt{s}(E_{+}+E_{-})-2E_{+}E_{-}} 
	\label{mselec}
\endequation
\equation
	m_{\chi_1} = \sqrt{m_{\se}^2-2m_{\se}
		\sqrt{E_{+}E_{-}}} .
	\label{mchi1}
\endequation
The cross section for the above mentioned  selectron production and decay 
is given by
\equation
	\sigma\left( e_R^-\gamma\rightarrow e^-
		\chi_1\chi_1\right)
		= |N^{'}_{11}-N^{'}_{12}{\rm tan}\theta_W|^4
		\sigma\left( e_R^-\gamma\rightarrow e^-
                \widetilde{\gamma}\widetilde{\gamma}\right) ,
	\label{eq:relxsects}
\endequation
where $\theta_W$ is the weak mixing angle.
The cross section on the right hand side is obtained in the limit when
the photino is the LSP.
Therefore, a measurement of the electron energy spectrum (and the total cross 
section) can determine $m_{\se}$, $m_{\chi_1}$ and 
$|N^{'}_{11}-N^{'}_{12}{\rm tan}\theta_W|$.
 Since $m_{\chi_1}$ and $N^{'}_{11}-N^{'}_{12}{\rm tan}\theta_W$
are functions of  $\mu$, $\tan\beta$, $M_1$ and $M_2$ only,
nontrivial constraints on these four parameters can be obtained.

  If $\chi_1$ is higgsino-like and $\chi_2$ is gaugino-like, the reaction
may proceed through the gaugino component 
of $\chi_2$, $e_R \gamma \rightarrow \ser \chi_2$, with subsequent decays of
the selectron  ($\ser \rightarrow e \chi_2$, where $\chi_2$ can be
real or virtual\footnote{Note that there is no coupling of the right-handed
selectron to charginos in the massless electron limit.}) and $\chi_2$. 
The electron from $\ser$ decay again has a flat distribution whose
endpoints determine $m_{\se}$ and $m_{\chi_2}$. 
The experimental signature depends on the decay modes of $\chi_2$.
For example, the radiative decay $\chi_2 \rightarrow \chi_1 \gamma$
can be the dominant mode if $\chi_1$ is higgsino-like and 
$\chi_2$ is gaugino-like, and the experimental signature 
is then $e\gamma\gamma + \eslash$. 
This signature also arises when the gravitino is the LSP
and it will be discussed in detail later in Section III. 
Cuts on the invariant mass of the $\eslash$ around $m_Z$ will 
get rid of the SM background.

\subsection{Double Step Function Behavior of the Outgoing Electron Energy 
Spectrum}  
 
  If both $\chi_1$ and $\chi_2$ contain non-negligible gaugino components,
their couplings to the electron and selectron will not be suppressed.
  Consider first the case in which $\chi_2$ is above the production 
threshold but is lighter than the $\ser$,  
and $\chi_{3,4}$ are heavier than the $\ser$ or are higgsino-like.
 Then we have
\begin{equation}
e_R \gamma \rightarrow \ser \chi_1 \;\;\;\;\;\;\;\;\;\;
\mbox{and} \;\;\;\;\;\;\;\;\;\;
\ser \rightarrow e \chi_1/e \chi_2  \label{eqx1x2}.
\end{equation}
  The electron energy spectrum is now the superposition of two step functions
with their endpoints in the $e\gamma$ CM frame $E_{\pm}^i$ ($i=1,2$) given by
\begin{eqnarray}
E_{\pm}^i & = & \frac{\sqrt{s}}{4} \left(1-
	\frac{m_{\chi_i}^2}{m_{\se}^2}\right)
    \left(1+ \frac{m_{\se}^2 - m_{\chi_1}^2}{s}\right)
    ( 1 \pm \overline{\beta}_{\se}),
\end{eqnarray}
where $\overline{\beta}_{\se}$ is the velocity of the $\ser$,
$\overline{\beta}_{\se}=\lambda^{\frac{1}{2}}(s,m_{\se}^2,m_{\chi_1}^2)/
(s + m_{\se}^2 - m_{\chi_1}^2)$ and $\lambda(x,y,z) \equiv x^2+y^2+z^2
-2xy-2yz-2xz$.

  The cross section for the process in Eq.~(\ref{eqx1x2}) is 
\begin{eqnarray}
 \sigma (e_R \gamma \rightarrow e X) 
   & = & \sigma (e_R \gamma \rightarrow \ser \chi_1)
 [ \mbox{BR}(\ser \rightarrow e \chi_1) 
    + \mbox{BR}( \ser \rightarrow e \chi_2)] \nonumber \\
 & = & \sigma (e_R \gamma \rightarrow \ser \photino_1)
            |N^{'}_{11}-N^{'}_{12}{\rm tan}\theta_W|^2 \times  
      [ \mbox{BR}(\ser \rightarrow e \chi_1) 
    + \mbox{BR}( \ser \rightarrow e \chi_2)]  \label{eqn:dstep}
\end{eqnarray}
where $X=\chi_1 \chi_1, \; \chi_1 \chi_2$, and  
$\photino_1$ denotes a photino with mass
 $m_{\chi_1}$. The branching ratios are given by
\equation
	\mbox{BR}(\ser \rightarrow e \chi_i) = \frac{
 |N^{'}_{i1}-N^{'}_{i2}{\rm tan}\theta_W|^2
		\left(1 - m_{\chi_i}^2/m_{\se}^2\right)^2}{
\sum_{j=1,2} |N^{'}_{j1}-N^{'}_{j2}{\rm tan}\theta_W|^2
		\left(1 - m_{\chi_j}^2/m_{\se}^2\right)^2}
	\;\;\;\;i=1,2.
	\label{branch12}
\endequation
The electron energy spectrum is plotted in 
Fig.~\ref{fig:dstep} for some representative masses.

  Besides the signature $e+\eslash$ from the chain 
$e_R\gamma \rightarrow \ser\chi_1 \rightarrow e \chi_1 \chi_1$,
the experimental signature also depends on the $\chi_2$ decay modes from 
the chain $e_R\gamma \rightarrow \ser\chi_1 \rightarrow e \chi_1 \chi_2$.
The signatures that we will focus on are $eX+\eslash$ where
$X$ contains no electrons and thus
measurement of the single electron energy spectrum can be easily performed. 
These include for example, the following $\chi_2$ decay modes:
i) the invisible decay  $\chi_2 \rightarrow \chi_1 \nu\bar{\nu}$
that gives rise to the signature $e + \eslash$; 
ii) the radiative decay $\chi_2 \rightarrow \chi_1 \gamma$ which leads to
 the signature  $e \gamma + \eslash$; 
and iii) the decay through the lightest Higgs boson,
$\chi_2 \rightarrow \chi_1 h \rightarrow \chi_1 b \bar{b}$, with the signature
$ej_bj_b + \eslash$, where $j_b$ denotes a $b$-jet.  If $m_{\chi_2}>m_h$
then the $b$-quarks will reconstruct the Higgs.
A comprehensive study of the various decay modes of $\chi_2$ can be
found in Ref.~\cite{neutralino}.
In the first two cases, the SM backgrounds involve escaped neutrino pairs
 from $Z$ decay and can be effectively  eliminated by imposing cuts on 
the invariant mass of the $\eslash$ around $m_Z$.  In the third case
the $\eslash$ and/or the two $b$-jets should reconstruct the $Z$ for
the SM background.

  The  masses of the selectrons and neutralinos, $m_{\se}$, $m_{\chi_1}$ and 
$m_{\chi_2}$, can be determined once the endpoints of the two step functions
 $E_{\pm}^i$ ($i=1,2$) are measured from the electron energy spectrum
 (see Fig.~\ref{fig:dstep} for illustration).
Furthermore, measurement of the partial cross section
for $e_R\gamma \rightarrow e\chi_1\chi_1$ from one 
of the two step functions fixes a relation between
$|N^{'}_{11}-N^{'}_{12}{\rm tan}\theta_W|$ and
$|N^{'}_{21}-N^{'}_{22}{\rm tan}\theta_W|$.  
Measurement of the other step function gives a lower bound on the
partial cross section for $e_R\gamma \rightarrow e\chi_1\chi_2$
and thus lower bounds on both 
$|N^{'}_{11}-N^{'}_{12}{\rm tan}\theta_W|$ and
$|N^{'}_{21}-N^{'}_{22}{\rm tan}\theta_W|$.
If $\chi_2$ decays only via the above three modes,
then both $|N^{'}_{11}-N^{'}_{12}{\rm tan}\theta_W|$ and
 $|N^{'}_{21}-N^{'}_{22}{\rm tan}\theta_W|$ can  be determined.   
Recall that $m_{\chi_1}$, $m_{\chi_2}$ and $N^{'}_{ij}$ are functions of 
$\mu$, $\tan \beta$, $M_1$ and $M_2$.
These latter four parameters
can thus be solved from the four known quantities,
$m_{\chi_1}$, $m_{\chi_2}$, $|N^{'}_{11}-N^{'}_{12}{\rm tan}\theta_W|$
and $|N^{'}_{21}-N^{'}_{22}{\rm tan}\theta_W|$.

   The other possibilities, including the case
 $e_R\gamma \rightarrow \ser \chi_1/\chi_2$ with
$\ser \rightarrow e \chi_1/\chi_2$ where the outgoing electron energy 
spectrum is a superposition of four step functions, can be
similarly analyzed. 
The experimental signatures are in general more involved, and no further
information about the MSSM can be gained beyond $m_{\se}$, 
$\mu$, $\tan \beta$, $M_1$ and $M_2$.

\section{Gravitino As LSP}

  Models of low energy gauge-mediated supersymmetry breaking \cite{dine}
have quite a different mass spectrum from the MSSM.
In particular, it is very likely that this SUSY breaking scheme would lead to
the gravitino being the LSP with the NLSP being either the 
lightest neutralino or a right-handed slepton, which we will assume to be
$\ser$.  
The gravitino gets its mass via the super Higgs mechanism
(assuming zero cosmological constant) with
$m_{\widetilde{G}}=\kappa d/\sqrt{6} \simeq 
1.7 \,(\sqrt{d}/100\; \mbox{TeV})^2 \; \mbox{eV}$, where 
$d$ denotes the scale of supersymmetry breaking and
$\kappa=\sqrt{8\pi G_{Newton}}$. 
In models of low energy dynamical supersymmetry breaking (DSB), 
  $\sqrt{d}$ ranges from  $100 \; \mbox{TeV}$ to  a few thousand TeV,  
and the gravitino mass takes its values in the eV to keV range.
 
  It is well known \cite{Fayet} that a light gravitino
can couple to matter with weak interaction strength instead of gravitationally
via its Goldstino component.
 The Goldstino has couplings to a particle and its superpartner determined
by the supersymmetric analog of the Goldberg-Treiman relation,
and is proportional to 
$ \Delta m /2d$, where $\Delta m$ measures the mass splitting between
an ordinary particle and its superpartner. 
This allows the NLSP (or any heavier sparticle) to decay into its SM 
partner and a gravitino with universal coupling strength. 

   Consider first the case of the lightest neutralino $\chi_1$ being the
NLSP. 
Since the cross section for direct production of a higgsino-like $\chi_1$ 
 via $e\gamma$ collision is negligible due to the tiny Yukawa coupling of the
electron, we will restrict our analysis to a gaugino-like $\chi_1$ and
focus on the dominant process,  
\begin{equation}
e_R \gamma \rightarrow \ser \chi_1 \rightarrow  e \chi_1 \chi_1  
\;\;\;\;\;\;\;\;\;\;\;\;\mbox{and}\;\;\;\;\;\;\;\;\;\;\;\;
 \chi_1 \rightarrow \gamma \widetilde{G},
\end{equation}
where $\widetilde{G}$ denotes the gravitino.
If $m_{\chi_1} > m_Z$, the decay $\chi_1 \rightarrow Z \widetilde{G}$ can
proceed through the zino component of $\chi_1$; 
For most of the parameter space,
the photonic decay will be the dominant mode, with its width given by
\cite{Fayet}
\begin{equation}
\Gamma(\chi_1 \rightarrow \gamma \widetilde{G}) =
|N^{'}_{11}|^2 \, m_{\chi_1}^5\,/(8 \pi d^2)
\simeq 1.1 \times 10^{-2}\,|N^{'}_{11}|^2\, (m_{\chi_1}/100\; \mbox{GeV})^5
\,(1 \; \mbox{eV}/m_{\widetilde{G}})^2 \; \mbox{eV} ,
\end{equation}
where $N^{'}_{11}$ is defined in Eq.~(\ref{eq:N}). 
 This translates into a decay distance for $\chi_1$ given by
\begin{equation}
D(\chi_1 \rightarrow \gamma \widetilde{G})
 \simeq 1.8 \times 10^{-3} \, |N^{'}_{11}|^{-2}\,
 \sqrt{E_{\chi_1}^2/m_{\chi_1}^2 -1}\, 
(m_{\widetilde{G}}/1 \; \mbox{eV})^2\,
(100\; \mbox{GeV}/m_{\chi_1})^5\; \mbox{cm} \label{eqdecaydist} ,
\end{equation}
where $E_{\chi_1}$ is the energy of $\chi_1$. 
For $m_{\widetilde{G}} \sim \mbox{eV} \; \mbox{--} \; \mbox{keV}$ and
$m_{\chi_1} \sim 100 \mbox{GeV}$, the decay length of $\chi_1$
ranges from hundreds of microns to tens of meters \cite{gravitinoph}.
It is therefore possible to observe the background-free signature
``$e +\;\mbox{displaced}\; \gamma\gamma + \eslash$''.
The electron energy spectrum has the characteristic
flat distribution.  If the supersymmetry breaking scale $\sqrt{d}$ is
several thousand TeV then the
decay will occur outside the detector.  In this case the signature
would be $e+\eslash$ and would be indistinguishable from the neutralino
scenario considered in Section II.

  If the NLSP decays quickly, the experimental signature will be
$e\gamma\gamma + \eslash$ and  has the SM background from 
$e_R\gamma \rightarrow e\gamma\gamma Z$ with $Z \rightarrow \nu \bar{\nu}$.
Cuts on the invariant mass of the $\eslash$ at $m_Z$
should again allow one to remove the SM background.

   The signature $e\gamma\gamma + \eslash$ can also arise in the MSSM
where $e_R \gamma \rightarrow \ser \chi_2$
with $\ser \rightarrow e \chi_2$ and $\chi_2 \rightarrow \chi_1 \gamma$.
This happens if $\chi_1$ is higgsino-like, $\chi_2$ is gaugino-like,
 and $\chi_2$ has a large branching ratio into $\chi_1$ and $\gamma$. 
Examination of the neutralino mass matrix \cite{neutralino,kane} shows 
this could indeed occur for certain ranges of parameter space 
(for example, $\tan \beta \simeq 1$ and 
$-\mu=|\mu|< M_1\simeq M_2$, as noted in ref.~\cite{kane}).
The electron energy spectrum is in both scenarios a step function 
whose endpoints allow for a determination of $m_{\se}$ and $m_{\chi_1}$
($m_{\chi_2}$) in DSB (MSSM) (cf. Eqs.~(\ref{mselec}) and (\ref{mchi1})).
The endpoints in the photon energy spectrum can then be used
to determine the LSP mass and therefore to distinguish unambiguously 
between the DSB and MSSM scenarios. 
We now derive the photon energy distributions in both scenarios, 
denoting the LSP and NLSP by $X_1$ and $X_2$ for convenience 
of presentation.

  Consider first the reaction chain $e_R \gamma \rightarrow \ser X_2$
with $X_2 \rightarrow X_1 \gamma$ 
\footnote{We are assuming for simplicity that the branching ratios
for $X_2 \rightarrow X_1 \gamma$ and $\ser \rightarrow e X_2$
are both equal to one.  The endpoints of the electron and photon
energy spectra, which are used to
distinguish between the MSSM and DSB scenarios, are independent
of this assumption.}.
The photon distribution in the rest frame of the $X_2$ is isotropic.  
Boosting back to the lab frame ($e\gamma$ CM frame) 
gives a flat energy distribution for the photon,
\begin{eqnarray}
\frac{d\sigma_1}{dE} & = & \frac{\sigma(e_R\gamma \rightarrow \ser X_2)}
{E_1^+ - E_1^-}  \;\;\;\;\;\;\; \mbox{for $E_1^- \le E \le E_1^+$},
\end{eqnarray}
where $E_1^{+}$ and  $E_1^{-}$ are the maximum and minimum photon energies
in the lab frame,
\begin{eqnarray}
	E_1^{\pm} & = & \frac{\sqrt{s}}{4}
		\left(1-\frac{m_{X_1}^2}{m_{X_2}^2}\right)
		\left(1 - \frac{m_{\se}^2 - m_{X_2}^2}{s}\right)
		\left( 1 \pm \beta_{X_2}\right), \label{eqde1}
\end{eqnarray}
and where $\beta_{X_2}=\lambda^{\frac{1}{2}}(s,m_{\se}^2,m_{X_2}^2)/
(s - m_{\se}^2 + m_{X_2}^2)$ is the velocity of $X_2$ in the lab frame. 

  We now turn to the second reaction chain, $e_R \gamma \rightarrow \ser X_2$
with the subsequent decays $\ser \rightarrow e X_2$ and
$X_2 \rightarrow X_1 \gamma$.
The photon energy distribution in the lab frame can be obtained by  
boosting back first to the $\ser$ rest frame from the $X_2$ rest frame
followed by a boost into the lab frame, and it is given by
\begin{eqnarray}
\frac{d\sigma_2}{dE} & = & 
\frac{\sigma(e_R\gamma \rightarrow \ser X_2)}
{E_2^- + E_2^+ - E_2^{\alpha} - E_2^{\beta}} \times
\left\{ \begin{array}{ll}
   \ln E/E_2^- \;\;\;\;\; & \mbox{if $E_2^- \le E \le E_2^{\alpha}$} \\
  \ln \frac{1+\beta}{1-\beta} & \mbox{if $E_2^{\alpha} \le E \le E_2^{\beta}$}
  \\
   \ln E_2^+/E & \mbox{if $E_2^{\beta} \le E \le E_2^+$} \\
     0         & \mbox{otherwise}
     \end{array} 
     \right.       \label{eqde2}
\end{eqnarray} 
where
 $E_2^{\alpha}=\mbox{min}(E_2^{A}, E_2^{B})$, $E_2^{\beta}=\mbox{max}
(E_2^{A}, E_2^{B})$,
and 
\begin{eqnarray}
	E_2^{\pm,A,B} & = & \frac{\sqrt{s}}{8}
		\left(1+\frac{m_{X_2}^2}{m_{\se}^2}\right)
		\left(1-\frac{m_{X_1}^2}{m_{X_2}^2}\right)
		\left(1+ \frac{m_{\se}^2 - m_{X_2}^2}{s}\right)
		\left(1\pm\beta^{\prime}_{X_2}\right)
		\left(1\pm\beta_{\se}\right),
\end{eqnarray} 
with $E_2^-$, $E_2^{A}$, $E_2^{B}$ and $E_2^+$ given by
the $--$, $-+$, $+-$ and $++$ combinations respectively;
and  $\beta=\mbox{min}(\beta^{\prime}_{X_2},\beta_{\se})$, where
$\beta^{\prime}_{X_2}=(m^2_{\se}-m^2_{X_2})/(m^2_{\se}+m^2_{X_2})$ 
is the $X_2$ velocity measured in the rest frame of the $\ser$
and $\beta_{\se}=\lambda^{\frac{1}{2}}(s,m_{\se}^2,m_{X_2}^2)/
(s + m_{\se}^2 - m_{X_2}^2)$ is the $\ser$ velocity in the lab frame.

 The single photon energy spectrum measured in a real experiment has
contributions from both photons in the above discussed reaction chains.
Since there is no correlation in the energies of the two decay photons,
the single photon energy spectrum is simply given by the average
$d\sigma/dE_{\gamma}=(d\sigma_1/dE + d\sigma_2/dE)/2$.
Fig.~\ref{fig:singlephot} gives the single photon  energy spectra
for both DSB and MSSM where the difference between
the two curves is due to the LSP mass. 

  Alternatively, the spectrum of the sum of the 
two photon energies can be measured and used to distinguish between 
the MSSM and  DSB scenarios.
The differential cross section can be derived as
\begin{eqnarray}
\frac{d\sigma}{d(E_1^{\gamma}+E_2^{\gamma})} & = & 
\frac{1}{E_1^+-E_1^-} \int_{E-E_1^+}^{E-E_1^-} H(x) dx
 \;\;\;\;\;\;\;\;\; \mbox{for $E_1^- + E_2^- \le E \le E_1^+ + E_2^+$},
\end{eqnarray}
where $H(E)\equiv d\sigma_2/dE$ as given by Eq.~(\ref{eqde2}).
These are plotted for the MSSM and DSB scenarios in Fig.~\ref{fig:sum}, 
where their difference
is essentially in the endpoints of the sum of the two photon energies
as a result of different LSP masses. 

  The analysis for $\chi_1 \rightarrow Z \widetilde{G}$ in DSB is similar 
to the discussion presented above for the photonic decay of $\chi_1$ and 
will not be repeated here.

  The second possibility in low energy DSB models is that  
the right-handed slepton plays the role of NLSP. 
We will assume this to be a $\ser$.
The decay chain of interest is
$e_R \gamma \rightarrow  \ser \chi_1$ with
$\chi_1 \rightarrow \ser e^+/\ser^{*}e^-$ and
$\ser (\ser^*) \rightarrow e^- (e^+) \widetilde{G}$.
The background-free signatures would be 
either three charged tracks without $\eslash$ if the selectrons
decay outside the detector,
 or ``$e^+ + \; \mbox{displaced}\; e^-e^- \; + \eslash$'' and 
 ``$e^- + \; \mbox{displaced}\; e^+e^- \; + \eslash$''
if the selectrons decay inside the detector.
For short decay lengths of the selectron,
the signature will be  $e^+e^-e^- + \eslash$.
The SM background in this case is quite involved, especially due
to the contributions from $W$'s, and
in general it cannot be eliminated by imposing 
cuts on the invariant mass of $\eslash$.  It appears that 
it may in fact be difficult to observe this signal over the 
background.
 
\section{Conclusion}

  Within the frameworks of the minimal supersymmetric standard model 
and low energy dynamical supersymmetry breaking,  
  we have explored the unique signatures arising from  high energy 
$e_R\gamma$ collisions,
$e+\eslash$ and $eX+\eslash$ ($X=\gamma, \gamma\gamma, j_bj_b, e^+e^-, 
\cdots$).
In cases where there is only one charged particle (the electron) in the
final state, the electron energy spectrum enjoys a flat distribution
or superposition of flat distributions. 

  Using a right-handed electron beam not only serves to eliminate the large
$W$ background, but also offers an efficient way to separate out
the remaining SM backgrounds by cuts on the 
invariant mass of the $\eslash$
carried away by a neutrino pair from $Z$ decay. 
The analysis is further simplified by the fact that 
only right-handed selectrons
will be produced and that these cannot decay into charginos.

In the MSSM, where the lightest neutralino $\chi_1$ is assumed to be the LSP, 
we have observed that a double step function behavior of 
the electron energy spectrum
may allow for a complete determination of the selectron mass and 
four basic parameters of the MSSM $\mu$, $\tan \beta$, $M_1$ and $M_2$,
and therefore provides an independent check of 
relations among these parameters
derived from higher scale physics like GUT.
  In low energy DSB, where the gravitino is the LSP, the NLSP
can be either the lightest neutralino or a right-handed slepton.
Depending on the supersymmetry breaking scale, the NLSP could have
long decay lengths and therefore could give rise to background-free 
signatures, 
$e+\;\mbox{displaced}\; \gamma\gamma + \eslash$ for the former,
and three charged tracks or $e^{\pm}+ \;\mbox{displaced} \;e^{\mp} e^- + 
\eslash$
for the latter.
Even when the dynamics is such that background-free decays do not
occur, we find that the outgoing electron and photon energy spectra
can be very useful diagnostic tools for distinguishing between the 
DSB and MSSM scenarios. 

\vskip 0.2in

We are grateful to Dr. G. Couture for allowing
us to use his Monte Carlo program which is useful for numerical checks
of our calculations.  K. K. would also like to thank N. Weiss for
helpful conversations.  
 This work is partially supported by the Natural Sciences and Engineering
Research Council of Canada.

\newpage

\noindent{\Large\bf Figure Captions}

\bigskip

\noindent
\refstepcounter{figcount}
{\bf Figure \thefigcount}:  (a) The ``double step function'' for
the outgoing electron energy spectrum in the MSSM (cf. Eq. (\ref{eqx1x2})).
The kinematics is such that $\chi_2$
cannot be produced initially, but may be
produced from the decay of the selectron.  For the purpose of 
illustration we have taken the LSP and NLSP to be the photino 
and zino, with masses 150 and 200 GeV, respectively,
and have set $\sqrt{s}=500$ GeV and $m_{\se}=300$ GeV.
Under these assumptions the 
selectron can only decay into the photino and zino
modes, with branching ratios 86\% and 14\%,
respectively.
The solid line shows the sum of the two step functions and the dotted
and dashed lines show the individual step functions due to photino and zino
production.  (b) The double step function for the same set 
of parameters as in (a) except that now
$m_{\chi_2}$$=$$250$ GeV.  In this case the branching ratios for 
the photino and zino modes are 95\% and 5\%.
\label{fig:dstep}

\smallskip
\noindent
\refstepcounter{figcount}
{\bf Figure \thefigcount}:  The single photon energy spectrum for
$e_R\gamma\rightarrow \ser X_2\rightarrow
eX_2X_2\rightarrow eX_1X_1\gamma\gamma$ where $X_1$ is the LSP
and $X_2$ is the NLSP.  In this
plot we have taken $X_2$ to be the photino and we have set
$\sqrt{s}$$=$$500$ GeV, $m_{\se}=200$ GeV and $m_{X_2}$$=$$100$
GeV.
The solid line corresponds to an LSP mass of 50 GeV (in the MSSM
scenario) and the dashed line corresponds to a massless LSP (which is
a good approximation in the gravitino scenario.)
  If $X_2$ is not pure photino then the solid (dashed) curve needs simply 
to be scaled by an overall factor of $|N_{21}^{\prime}-N_{22}^{\prime}
{\rm tan}\theta_W|^2$ ($|N_{11}^{\prime}-N_{12}^{\prime}
{\rm tan}\theta_W|^2$).
\label{fig:singlephot}

\smallskip
\noindent
\refstepcounter{figcount}
{\bf Figure \thefigcount}:  The differential spectrum for the sum 
of the two photon energies, $d\sigma/d(E_1^{\gamma}+E_2^{\gamma})$,
for the reaction 
$e_R\gamma \rightarrow \ser X_2 \rightarrow  eX_2X_2\rightarrow 
eX_1X_1\gamma\gamma$,
using the same parameters as in Fig.~\ref{fig:singlephot}.
\label{fig:sum}

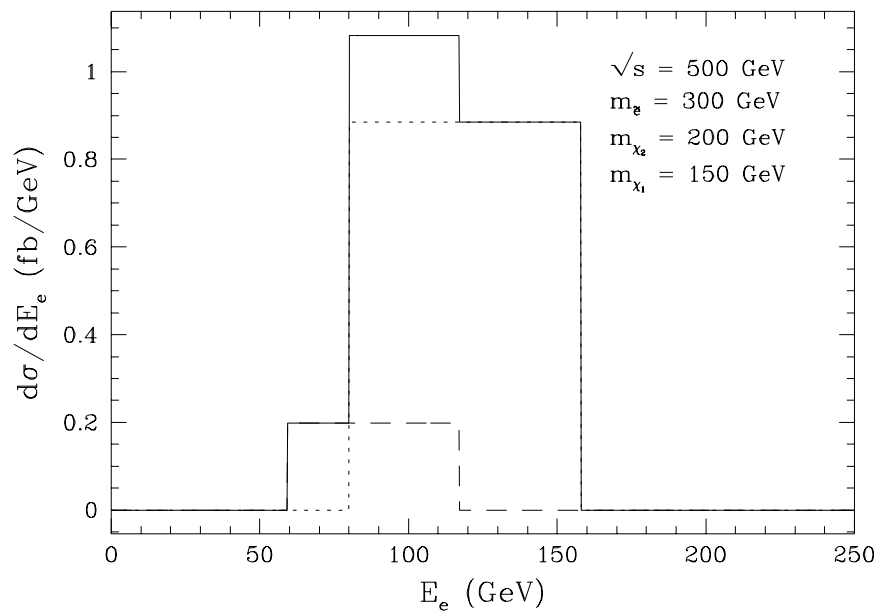
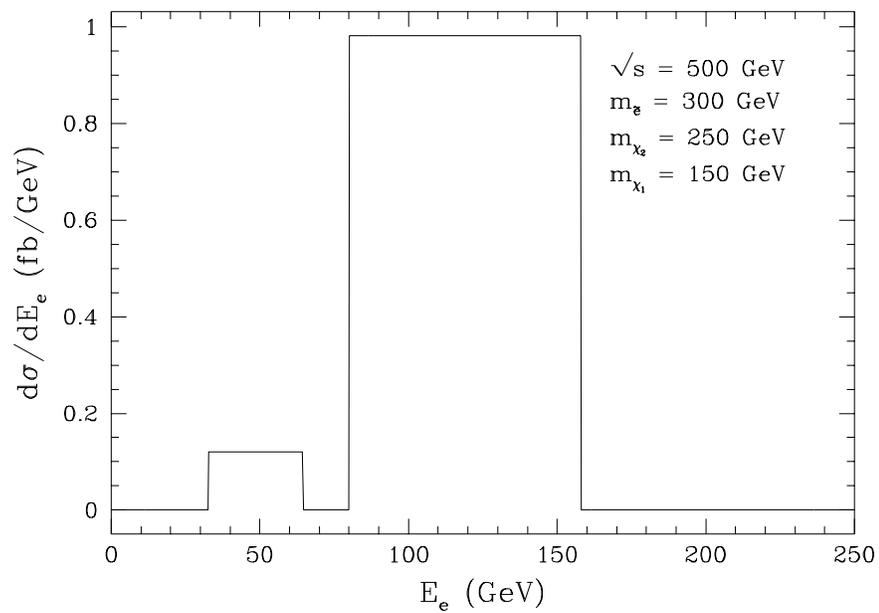
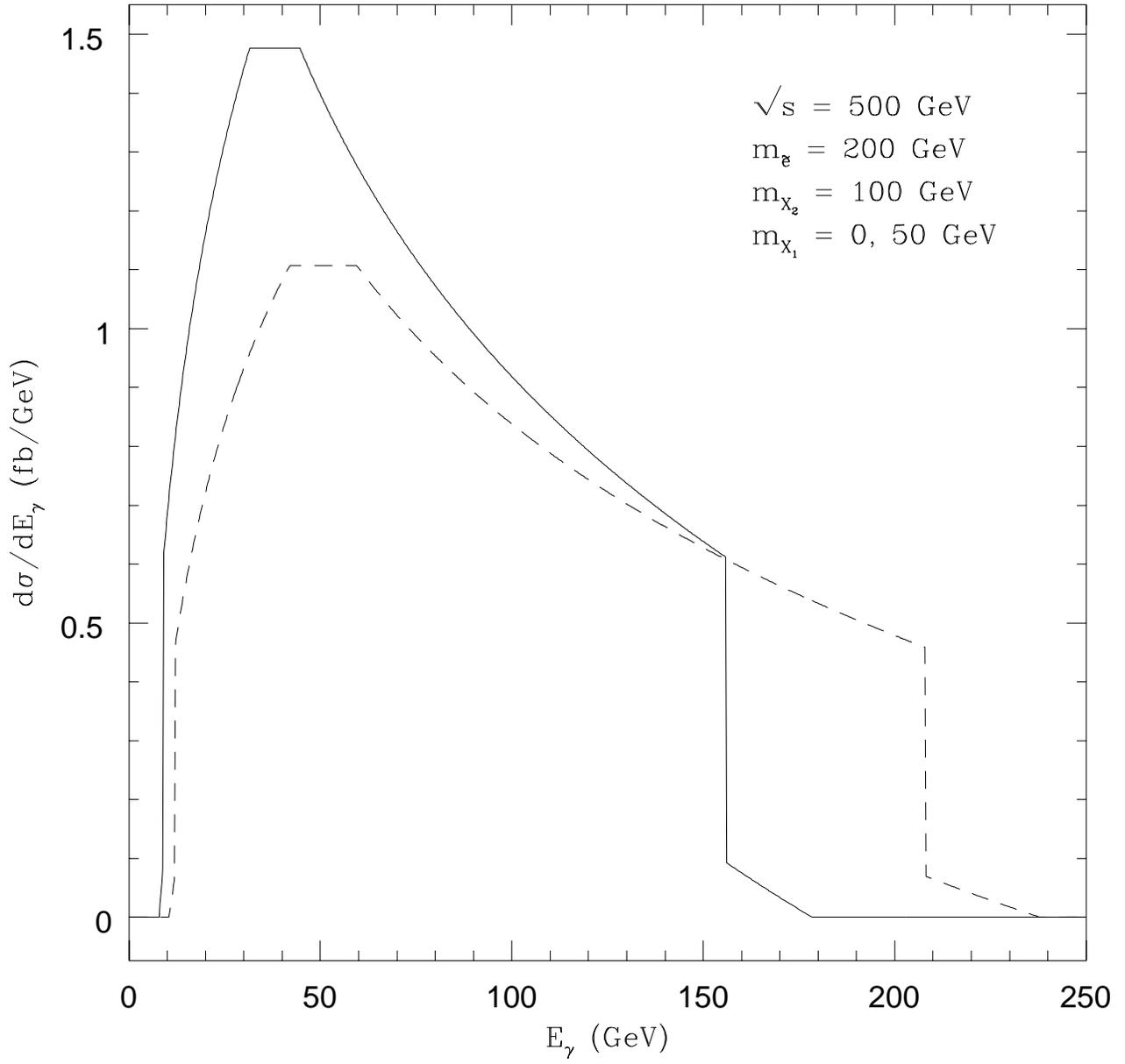
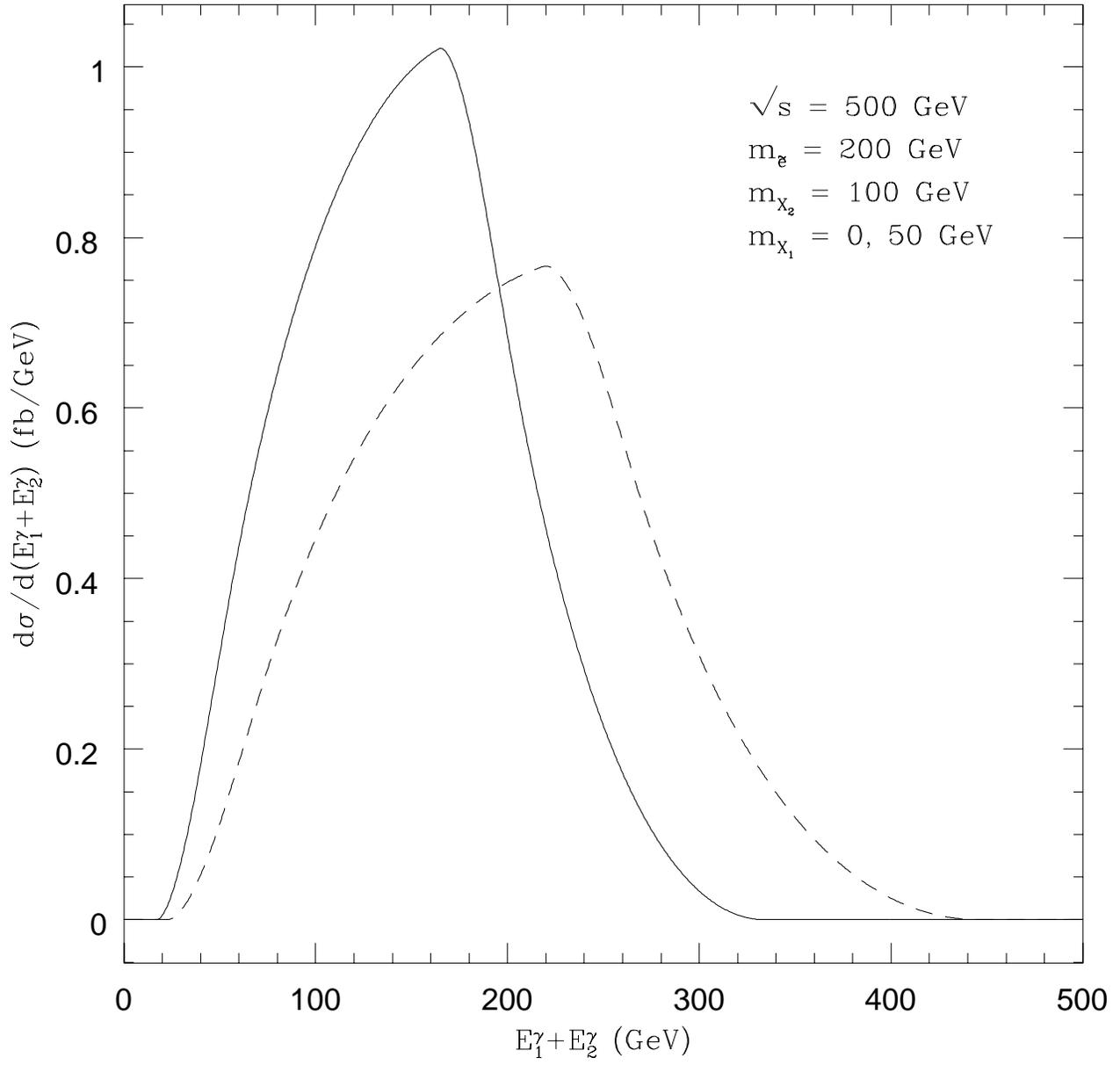
\end{document}